\documentclass[aps,amsmath,amssymb,prd,showpacs,floatfix,preprint,superscriptaddress,nofootinbib,12pt]{article}
\usepackage{jheppub}
\usepackage{bm}
\usepackage{ulem}

\usepackage{pstricks}
\usepackage{color}
\usepackage[utf8]{inputenc}

\newcommand{\p}{\partial}

\usepackage{epstopdf}

\allowdisplaybreaks[3]

\begin{document}

\title{\center{Under The Dome}\\
\center{\rm{Doped holographic superconductors with broken translational symmetry}}}

\author[a,b]{Matteo Baggioli,}
\author[c]{Mikhail Goykhman}
\affiliation[a]{Institut de F\'{i}sica d'Altes Energies (IFAE), Universitat Aut\`{o}noma de Barcelona,\\ The Barcelona Institute of
Science and Technology, Campus UAB,\\ 08193 Bellaterra (Barcelona), Spain}
\affiliation[b]{
Department of Physics, Institute for Condensed Matter Theory,
University of Illinois,\\
1110 W. Green Street, Urbana, IL 61801, USA}
\affiliation[c]{Enrico Fermi Institute, University of Chicago,\\
5620 S. Ellis Av., Chicago, IL 60637, USA}
\emailAdd{mbaggioli@ifae.es,goykhman@uchicago.edu}

\abstract{
We comment on a simple way to accommodate translational symmetry
breaking into the recently proposed holographic model which features a superconducting dome-shaped region on the temperature-doping phase diagram.
}

\maketitle

\section{Introduction}\label{sec:intro}

The original holographic model of \cite{Hartnoll:2008vx,Hartnoll:2008kx}
successfully describes the physics of superconducting (SC) phase transitions within a strongly coupled regime.
This is achieved by introducing a charged scalar field into the
finite-temperature, finite-density
AdS-RN black-brane setting.
The charged scalar field is stated to be the bulk dual of the
condensate of the boundary charge carriers and it represents indeed the natural order parameter for the SC phase transition.
When the system is in the normal phase, the bulk scalar is identically trivial; on the contrary in the SC phase, the bulk scalar develops a non-trivial profile.
This model however describes just one face of a variety of phenomena
exhibited by real-world materials. For the AdS/CMT field
to be a practically-oriented endeavor, one should augment the framework
of \cite{Hartnoll:2008vx,Hartnoll:2008kx} by coupling it to additional sectors,
with the intention to account for more of a non-trivial experimentally observed physics.

The present work is motivated by the recent paper \cite{Kiritsis:2015hoa}, which follows in this direction
by building a model which exhibits
a normal, superconducting, anti-ferromagnetic and striped/checkerboard
phases on the doping-temperature plane.
It is interesting that the superconducting phase of \cite{Kiritsis:2015hoa}
appears within a dome-shaped region in the middle of the phase plane as in actual High-Tc superconductors.
However the normal phase of \cite{Kiritsis:2015hoa} possesses
an infinite DC conductivity, a property which it shares with the original
holographic superconductor \cite{Hartnoll:2008vx,Hartnoll:2008kx} and which does not allow it to be labeled as a proper metallic phase.
Unlike the infinite DC conductivity of the superconducting phase,
which is a result of the condensation of the charge carriers,
an infinite DC conductivity in the normal phase is a straightforward
consequence of the translational invariance of the boundary theory, which needs to be relaxed.

Translational symmetry breaking mechanisms in holographic models have recently received plenty of attention
in the literature, in order to mimic more realistic condensed matter situations. Breaking the bulk diffeomorphism invariance via introducing a graviton mass \cite{Vegh:2013sk} is an efficient way of achieving it.
Massive gravity theories can be formulated covariantly in terms of the Stuckelberg fields and such a construction results in a finite DC conductivity \cite{Andrade:2013gsa}. The latter model has been considered in combination with the holographic
superconductor setting in the recent literature,
which includes \cite{Andrade:2014xca,Kim:2015dna,Baggioli:2015zoa}.

In this note we point out that the recent model
\cite{Kiritsis:2015hoa}, featuring a holographic SC dome, can be further improved by coupling
it to a neutral scalar sector, governed by a general Lagrangian as in \cite{Baggioli:2014roa} and responsible for the breaking of translational symmetry.
The resulting holographic superconductor can be studied in spirit of \cite{Baggioli:2015zoa}.
Its normal phase possesses a finite DC conductivity.
Our results prove that the superconducting dome of \cite{Kiritsis:2015hoa}
continues to exist once the translational symmetry has been broken.

Using the non-linear model proposed in \cite{Baggioli:2014roa}
we are able to describe a normal phase with conductivity decreasing upon lowering of
the temperature. The resulting system exhibits three phases on the temperature-doping
plane: superconducting, normal metallic and normal pseudo-insulating.
The phases are essentially distinguished by the DC conductivity:
infinite in the superconducting phase, decreasing with temperature
in the metallic phase, and growing with temperature in the pseudo-insulating phase.

The rest of this paper is organized as follows. In the next section
we set up the model which we study in this paper.
We describe the normal phase in section \ref{sec:normalphase},
where we also construct the metal/pseudo-insulator phase
diagram on the temperature-doping plane, for the model
governed by a non-linear Lagrangian for the neutral scalars.
In section \ref{sec:instability}
we determine the critical temperature and doping for which the
normal phase becomes unstable towards the development of a scalar hair.
This signals a superconducting phase transition,
which we confirm in section \ref{sec:condensate}
by solving numerically for the whole background and calculating
the temperature dependence of the charge condensate v.e.v.
We discuss our results in \ref{sec:discussion}.
In appendix \ref{app:BackgroundEqs} we collect the equations
of motion for the whole background that used in the paper.

 \section{The system}\label{sec:model}
 
In this section we set up the holographic system which we will be studying in this paper.
We follow closely the conventions of \cite{Kiritsis:2015hoa}.
We consider the following bulk degrees of freedom: the metric $g_{\mu\nu}$,
two $U(1)$ gauge fields $A_\mu$, $B_\mu$, the complex scalar field
$\psi$, and two neutral scalars $\phi ^I$, $I=x,y$.
Here $x,y$ are spatial coordinates on the boundary. We will denote the radial bulk coordinate as $u$.
The boundary is located at $u=0$, the horizon is located
at $u=u_h$.

We want to describe a system of charge carriers,
coexisting with a media of impurities. The density of the charge
carriers is denoted by $\rho_A$ and is dual to the gauge field $A_\mu$ while the density of impurity  
$\rho_B$ is dual to the gauge field $B_\mu$.
The quantity
\begin{equation}
\label{dopingdefinition}
{\bf x}=\rho_B/\rho_A
\end{equation}
is called the doping parameter and represents the amount of charged impurities present in the system
\cite{Kiritsis:2015hoa}.

The boundary system exists in a superconducting phase when 
the Bose condensate is formed. The vacuum expectation value of the condensate is the order parameter for
the superconducting phase transition and it is described holographically via the complex scalar field, $\psi$.
The latter is charged w.r.t. the $U(1)_A$ gauge field \cite{Hartnoll:2008vx,Hartnoll:2008kx}.

We introduce explicit translational symmetry breaking into the system by coupling it to a sector
of neutral and massless scalars $\phi^I$ with spatial dependent sources $\phi ^I=\alpha \, x^I$, $I=x,y$  \cite{Andrade:2013gsa}.
In this paper we will be making use of a generalized action for those scalars introduced in \cite{Baggioli:2014roa}.

The total action of the model is written as:
\begin{align}
\label{totalaction}
S=\frac{1}{16\pi}\int d^4x \sqrt{-g}\left(R+\frac{6}{L^2}+{\cal L}_c+{\cal L}_s\right)
\end{align}
where we fixed the cosmological constant $\Lambda=-3/L^2$, and denoted the Lagrangian densities for the charged sector \cite{Kiritsis:2015hoa},
and the neutral scalar sector \cite{Baggioli:2014roa} as:
\begin{align}
{\cal L}_c&=-\frac{Z_A(\chi)}{4}A_{\mu\nu}A^{\mu\nu}
-\frac{Z_B(\chi)}{4}B_{\mu\nu}B^{\mu\nu}
-\frac{Z_{AB}(\chi)}{2}A_{\mu\nu}B^{\mu\nu}\label{chargesectorlagrangian}\\
&-\frac{1}{2}(\partial_\mu \chi)^2-H(\chi)(\partial_\mu \theta-q_AA_\mu-q_B B_\mu)^2
-V_{int}(\chi)\\
{\cal L}_n&=-2m^2 V(X)\,.
\end{align}
Here the $A_{\mu\nu}$ and $B_{\mu\nu}$ stand for the field strengths of the
gauge fields $A_\mu$ and $B_\mu$ respectively.
Following \cite{Kiritsis:2015hoa} we decomposed the charge scalar as $\psi =\chi e^{i\theta}$.
We also defined:
\begin{equation}
X=\frac{1}{2}g^{\mu\nu}\p_\mu\phi^I\p_\nu\phi^I\,.
\end{equation}
The most general black-brane ansatz we consider is:
\begin{align}
\label{generalansatz}
ds^2&=\frac{L^2}{u^2}\left(-f(u)e^{-\tau(u)}dt^2+dx^2+dy^2+\frac{du^2}{f(u)}\right)\,,\\
A_t&=A_t(u)\,,\qquad B_t=B_t(u)\,,\\
\chi&=\chi (u)\,,\qquad \theta\equiv 0\,,\\
\phi^x&=\alpha\, x\,,\qquad \phi^y=\alpha\, y\,.
\end{align}
The corresponding equations of motion are provided in appendix \ref{app:BackgroundEqs}.
The temperature of the black brane (\ref{generalansatz}) is given by:
\begin{equation}
T=-\frac{e^{-\frac{\tau(u_h)}{2}}f'(u_h)}{4\pi}\,.
\end{equation}
In the rest of the paper we will be considering:
\begin{align}
V_{int}(\chi)&=\frac{M^2\chi^2}{2}\,.
\end{align}
Solving the $\chi$ e.o.m. near the boundary $u=0$ one obtains $\chi(u)=C_- \, (u/L)^{3-\Delta}
+C_+\, (u/L)^\Delta $, where
$(ML)^2=\Delta (\Delta -3)$.
Here $C_-$ is the source term, which one demands to vanish, and
$C_+$ is the v.e.v. of the dual charge condensate operator, $C_+=\langle {\cal O}\rangle$.
The $\Delta$ is equal to the scaling dimension of the operator ${\cal O}$. Following \cite{Kiritsis:2015hoa} in this paper we
fix the scaling dimension to be $\Delta =5/2$.
 
\section{Normal phase}\label{sec:normalphase} 
\label{sec:normalphase}

In the normal phase the charge condensate
vanishes, and the charged scalar field is trivial, $\chi\equiv 0$. Solving the background equations of motion we obtain $\tau\equiv 0$, along with:
\begin{align}
\label{normalphase}
f(u)&=u^3\int _{u_h}^udy\,\frac{\rho_A^2(1+{\bf x}^2)\,y^4+4\,(mL)^2\, V(\alpha^2\, y^2)-12}{4y^4}\,,\\
\label{normalphase2}
A_t(u)&=\rho_A(u_h-u)\,,\hspace{0.5cm}B_t(u)=\rho_B(u_h-u)\,.
\end{align}
The temperature in the normal phase is given by:
\begin{equation}
\label{generaltemperature}
T=\frac{12-\rho_A^2(1+{\bf x}^2)\,u_h^4-4\,(mL)^2 \,V(\alpha^2\,u_h^2)}{16\pi u_h}\,.
\end{equation}
Using the membrane paradigm one can calculate analytically the DC conductivity in the normal phase \cite{Blake:2013bqa,Donos:2014cya}. Its value for a general neutral scalars Lagrangian $V$ is given by \cite{Baggioli:2014roa}:
\begin{equation}
\label{generalsigmaDC}
\sigma_{DC}=1+\frac{\rho_A^2u_h^2}{2\,m^2\,\alpha^2\,\dot V(u_h^2\alpha^2)}\,.
\end{equation}
In particular for the linear Lagrangian:
\begin{equation}
V(X)=\frac{1}{2\,m^2}\,X\,,\label{linearV}
\end{equation}
we recover the set-up of \cite{Andrade:2013gsa}.
Choosing a non-trivial $V(X)$ one can incorporate more interesting physics.
As it was pointed out in \cite{Baggioli:2014roa} for certain $V(X)$
one can realize both a \textit{pseudo-insulating}\footnote{A DC conductivity increasing with temperature  
is reminiscent of an insulating behavior, although it is ubiquitous in the
considered holographic model that the zero-temperature
conductivity is always non-vanishing, as pointed out recently in \cite{Grozdanov:2015qia}.} and a metallic phases, characterized by the following peculiar temperature dependences of the DC conductivity:
\begin{align}
\frac{d\sigma_{DC}}{dT}>0\,:\quad {\rm pseudo-insulating}\,,\qquad
\frac{d\sigma_{DC}}{dT}<0\,:\quad {\rm metallic}\,.
\end{align}
A transition between these two states happens at critical temperature $T_0$ which is determined by:
\begin{equation}
\label{dsigmaDCdT}
\frac{d\sigma_{DC}}{dT}=0\quad\Rightarrow\quad \dot V(u_h^2\,\alpha^2)-u_h^2\,\alpha^2\,
\ddot V(u_h^2\,\alpha^2)=0\,.
\end{equation}
Solving for the horizon radius $u_h$ in terms of the temperature (\ref{generaltemperature}),
and plugging it into (\ref{dsigmaDCdT}), we obtain the phase transition
line $T_0(m,\alpha,{\bf x})$.
\begin{figure}
\begin{center}
\includegraphics[width=.45\textwidth]{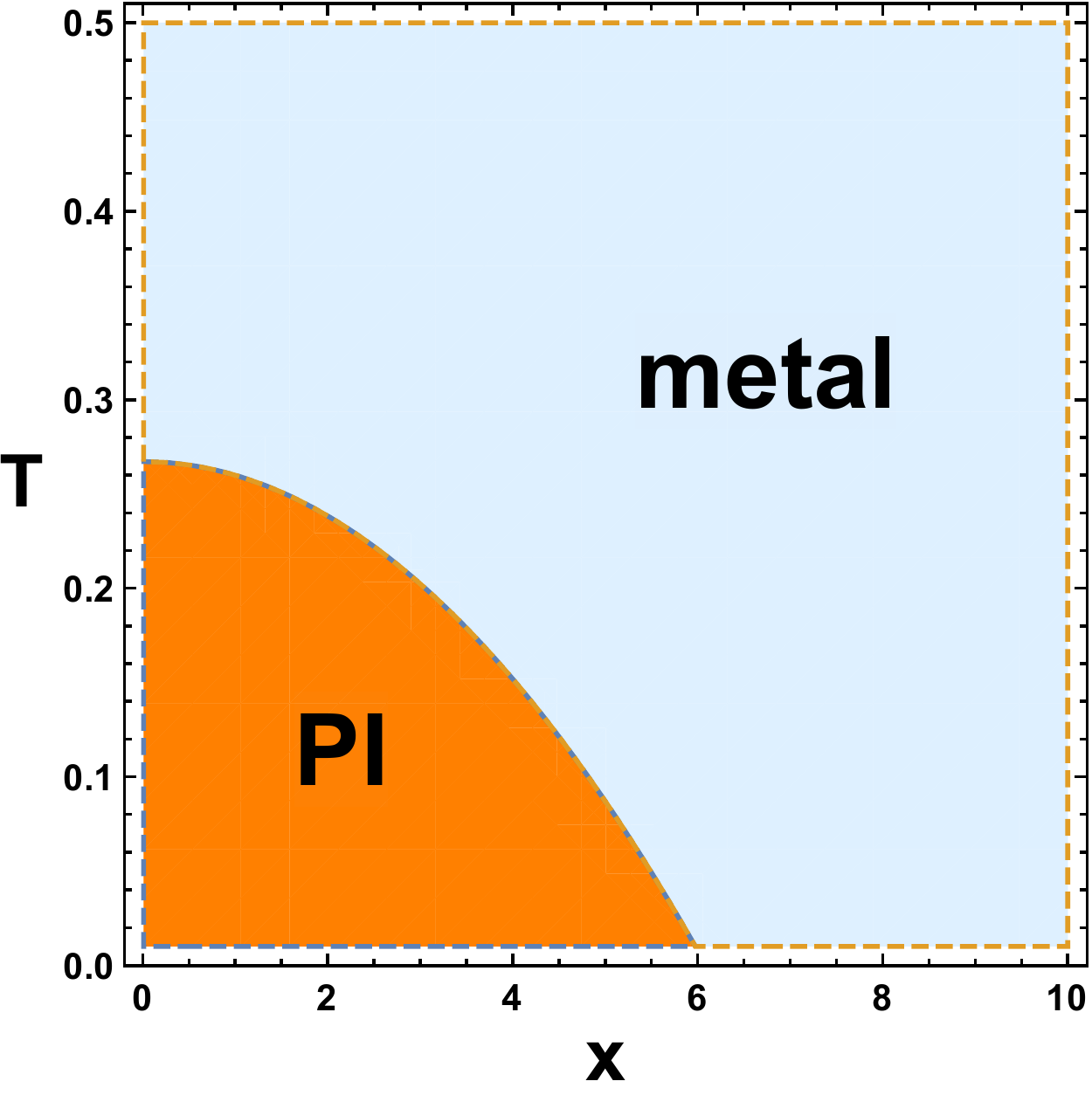}
\end{center}
\caption{Phase diagram of the model (\ref{nonlinearV}) with $m=1$, $\alpha =1$,
showing the pseudo-insulating phase, characterized by the DC conductivity
behavior $\sigma_{DC}'(T)>0$, and the metallic phase, characterized by the DC conductivity
behavior $\sigma_{DC}'(T)<0$.}
\label{fig:DCnonlinearV}
\end{figure}
Solution to (\ref{dsigmaDCdT}) exists for a non-trivial choice of the Lagrangian $V(X)$.
In this paper we will consider:
\begin{equation}
V(X)=X+X^5\,.\label{nonlinearV}
\end{equation}
For this $V(X)$, fixing ${\bf x}=0$, $\alpha={\cal O}(1)$,
one obtains a phase diagram on the $(m,T)$ plane, with the pseudo-insulating
phase occupying a compact corner region around the origin of the phase plane \cite{Baggioli:2014roa}.
This property has been embedded in the holographic superconductor phase diagram in \cite{Baggioli:2015zoa}.

In this paper, following \cite{Kiritsis:2015hoa}, we are interested in a phase structure
on the $({\bf x},T)$ doping-temperature plane. Therefore we fix $m$ and $\alpha$ and determine the critical
temperature $T_0({\bf x})$. For the model (\ref{nonlinearV})
it is possible to achieve a compact pseudo-insulating region around the origin of the $({\bf x},T)$ plane,
see figure \ref{fig:DCnonlinearV}.
 
\section{Instability}
\label{sec:instability}

To determine whether a boundary system exhibits a superconducting phase one can consider a normal phase of the bulk system and see whether it becomes unstable towards developing a non-trivial profile of the scalar $\chi (u)$.
This approach assumes that the corresponding superconducting phase transition is of the second order, which should be checked separately by solving the whole system away from the regime of a small $\chi$, which we do in section \ref{sec:condensate}. In this section we solve the linearized equation of motion for $\chi$ in the normal phase background.
Following \cite{Kiritsis:2015hoa} we define the following expansion of the couplings:
\begin{align}
\label{couplingexpansions}
&H(\chi)=\frac{n\,\chi^2}{2}\,,\hspace{0.3cm}
Z_A(\chi)=1+\frac{a\,\chi^2}{2}\,,\hspace{0.3cm}
Z_B(\chi)=1+\frac{b\, \chi^2}{2}\,,\hspace{0.3cm}
Z_{AB}(\chi)=\frac{c\, \chi^2}{2}\,.
\end{align}
and define the $U(1)_{A,B}$ charges to be $q_A=1\,,\,q_B=0$ .
In this section we use scaling symmetry and set the charge carriers density $\rho_A=1$, and express the impurity density in terms of the doping parameter $\rho_B={\bf x}$.

A natural place to start searching for superconductor
is at zero temperature.
When the temperature is zero, the infra-red limit of the bulk geometry is $AdS_2\times \mathbb{R}^2$, where the scale of the $AdS_2$ is given by:
\begin{equation}
L_2^2=\frac{2L^2}{u_h^2\,f''(u_h)}\,.
\end{equation}
The effective mass of the scalar $\chi$ can be read off from its linearized equation of motion and is given by:
\begin{equation}
M_{eff}^2=\frac{1}{2\,f''(u_h)}\,\left[f''(u_h)\left(2\,(ML)^2-(a+2\,c\,x+b\,x^2)\,u_h^4\right)-4\,n\,u_h^2\,(q_A+{\bf x}\, q_B)^2\right]\,.
\end{equation}
The system becomes unstable towards developing a non-trivial $\chi(u)$
profile if the BF bound for the scalar $\chi$ is violated in the $AdS_2$, namely $M_{eff}^2 L_2^2< -\frac{1}{4}\,$, or more specifically:
\begin{align}
(2\,M^2-u_0^4\,(a+2\,c\,x+b\,x^2))\,(6+m^2((\alpha\,u_0)^2\,\dot V-2V))-2\,n\,u_0^4\,(
q_A+q_B\,x)^2<0\,,\label{AdS2BFbound}
\end{align} 
where dot stands for derivative of $V$ w.r.t. its argument and $u_0$ for the radial position of the extremal horizon, $T(u_0)$=0. We also set $L=1$.

To obtain a superconducting dome on the temperature-doping plane $(T,{\bf x})$, one needs to fix the parameters of the model in such a way that zero-temperature superconducting instability appears in an interval $[{\bf x}_1,{\bf x}_2]$, between two positive values ${\bf x}_{1,2}$ of the doping parameter.
In the context of instability analyses tuning the model amounts to a choice of
the coefficients $a$, $b$, $c$, $n$, appearing in the expansion (\ref{couplingexpansions}).
In \cite{Kiritsis:2015hoa} the specific model determined by the parameters:
\begin{equation}
\label{Kiritsisparameters}
a=-10\,,\quad b=-\frac{4}{3}\,,\quad c=\frac{14}{3}\,,\quad n=1\,.
\end{equation}
has been extensively studied, and it was pointed out that in the
interval ${\bf x}\in [{\bf x}_1,{\bf x}_2]$, ${\bf x}_1\simeq 1.25$, ${\bf x}_2\simeq 5.8$ at zero temperature the effective mass of the scalar field $\chi$ violates the $AdS_2$ BF bound.

\begin{figure}
\begin{center}
\includegraphics[width=.45\textwidth]{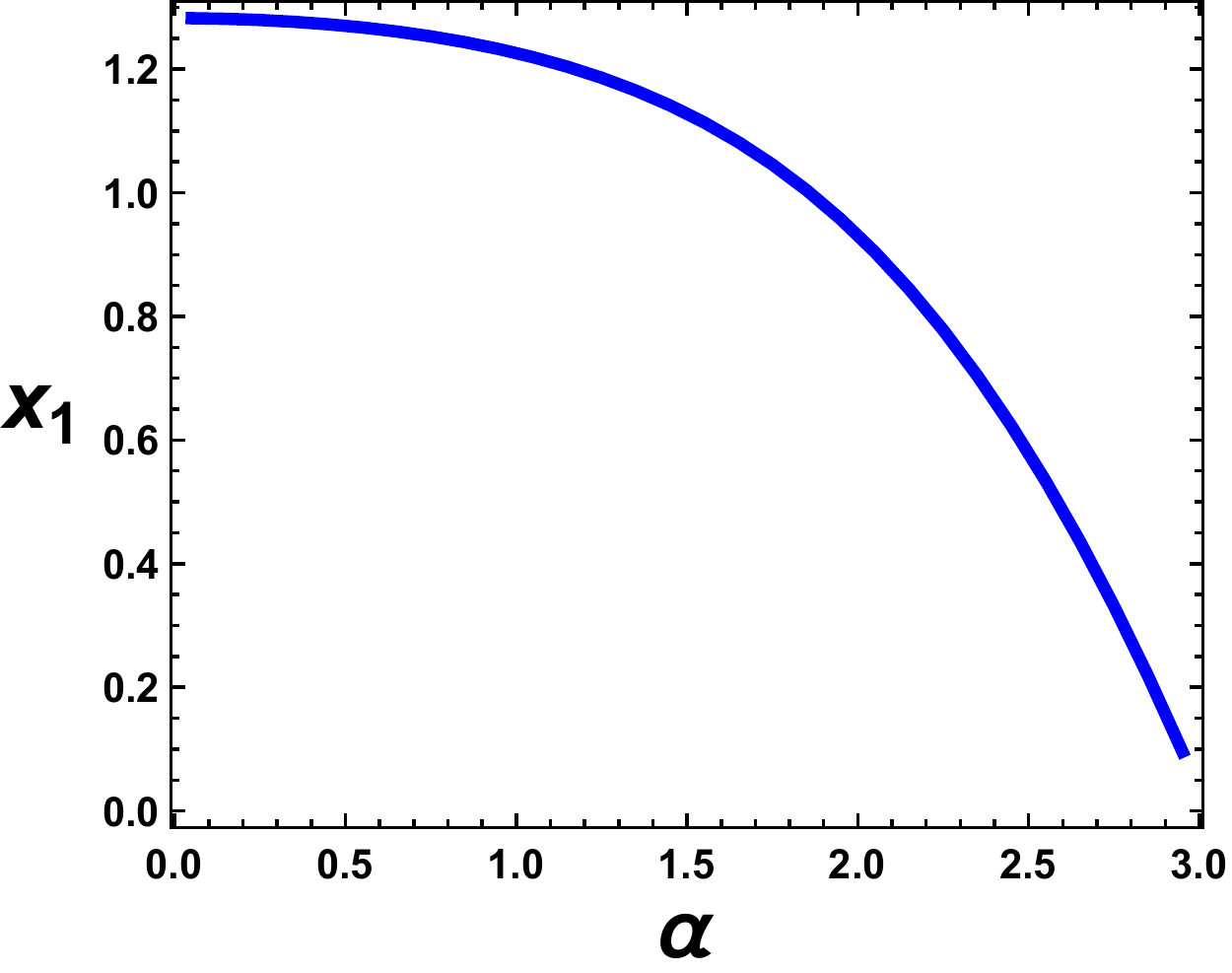}
\includegraphics[width=.45\textwidth]{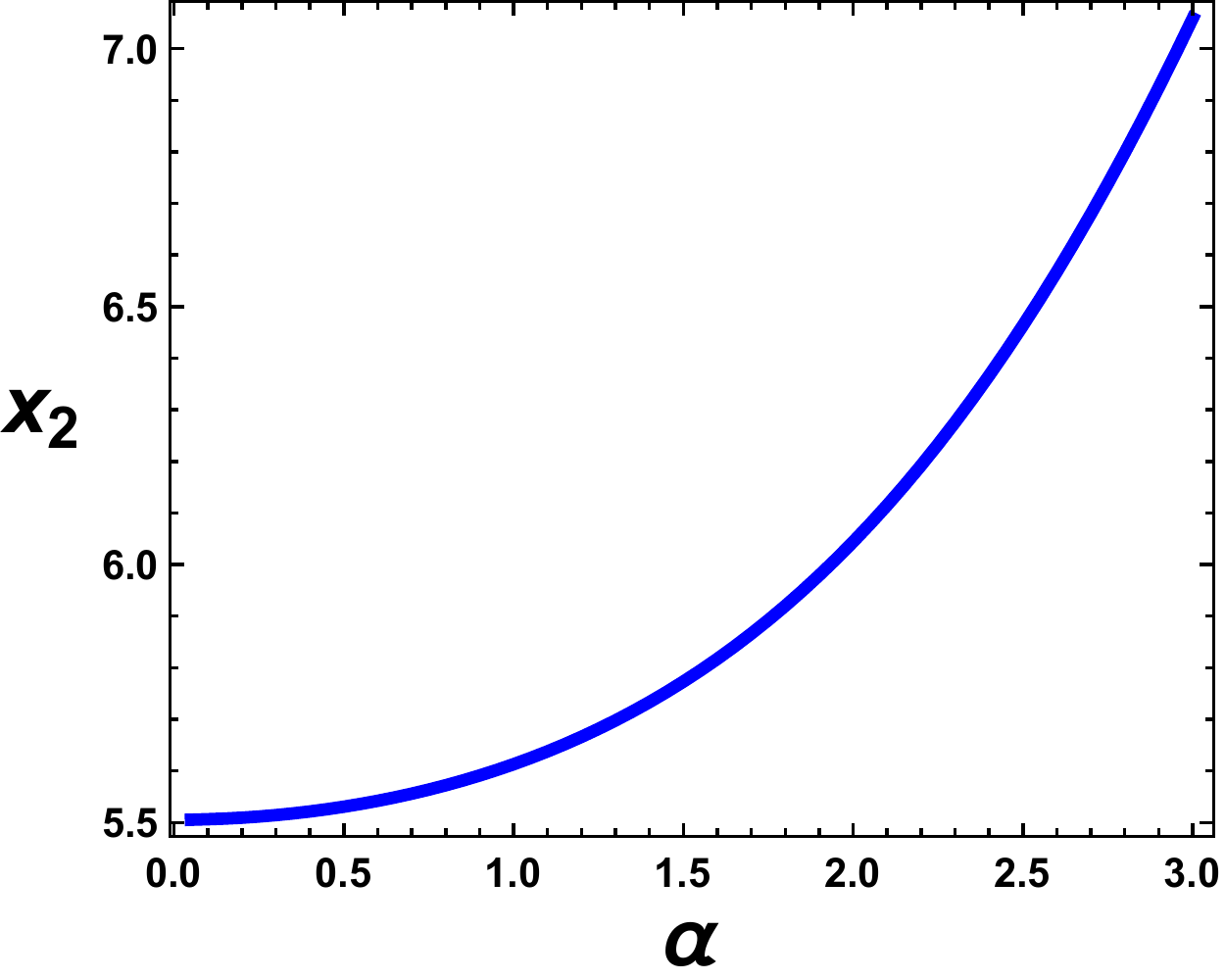}
\end{center}
\caption{The boundaries of the zero-temperature IR instability region 
on the doping line for the model (\ref{Kiritsisparameters}), with the
translational symmetry broken by the neutral scalars with the Lagrangian
(\ref{linearV}).}
\label{fig:x12ofalpha}
\end{figure}

Now let us consider the model (\ref{Kiritsisparameters})
but with the translational symmetry broken by the neutral scalars with the linear Lagrangian 
(\ref{linearV}).
We observe that for $\alpha\neq 0$ the instability persists, although
the `depth' of the $AdS_2$ BF violation becomes smaller, and therefore we expect the corresponding
critical temperature of the superconducting phase transition to be lower. This is as to say that the breaking of translational symmetry unfavores the SC instability.
We plot the $\alpha$-dependence of the boundary
points of the IR instability region, ${\bf x}_{1,2}(\alpha)$, in figure \ref{fig:x12ofalpha}.\\

\begin{figure}
\begin{center}
\includegraphics[width=.43\textwidth]{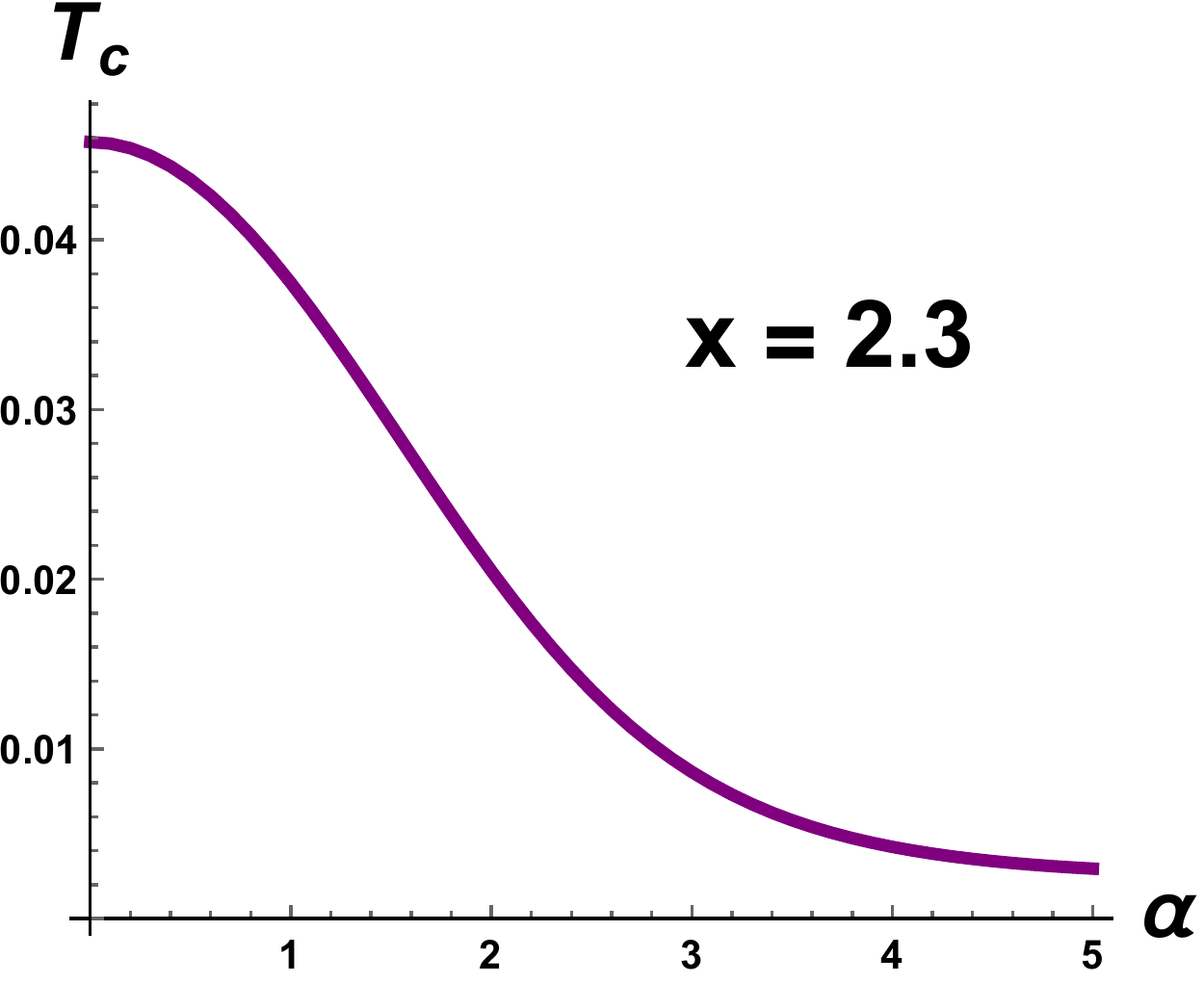}
\end{center}
\caption{Critical temperature $T_c(\alpha)$ for the model (\ref{Kiritsisparameters}), with the
translational symmetry broken by the neutral scalars with the Lagrangian
(\ref{linearV}).
Here the doping is fixed to be ${\bf x}=2.3$.}
\label{fig:TcMass}
\end{figure}

The critical temperature $T_c$ of a second-order phase transition can be determined by studying the dynamics of the scalar $\chi (u)$, considered as a probe in a finite-temperature normal phase background.
We are looking for a maximal value of the temperature for which the source coefficient $C_-$
of the near-boundary expansion of the field $\chi$ vanishes.
Consider the model with the linear Lagrangian (\ref{linearV}).
It is interesting to observe how the critical temperature depends on the magnitude
$\alpha$ of the translational symmetry breaking.
In accordance with our expectations from the zero-temperature instability
analyses we observe a decrease of the critical temperature with $\alpha$,
as shown in figure \ref{fig:TcMass}.

\begin{figure}
\begin{center}
\includegraphics[width=.55\textwidth]{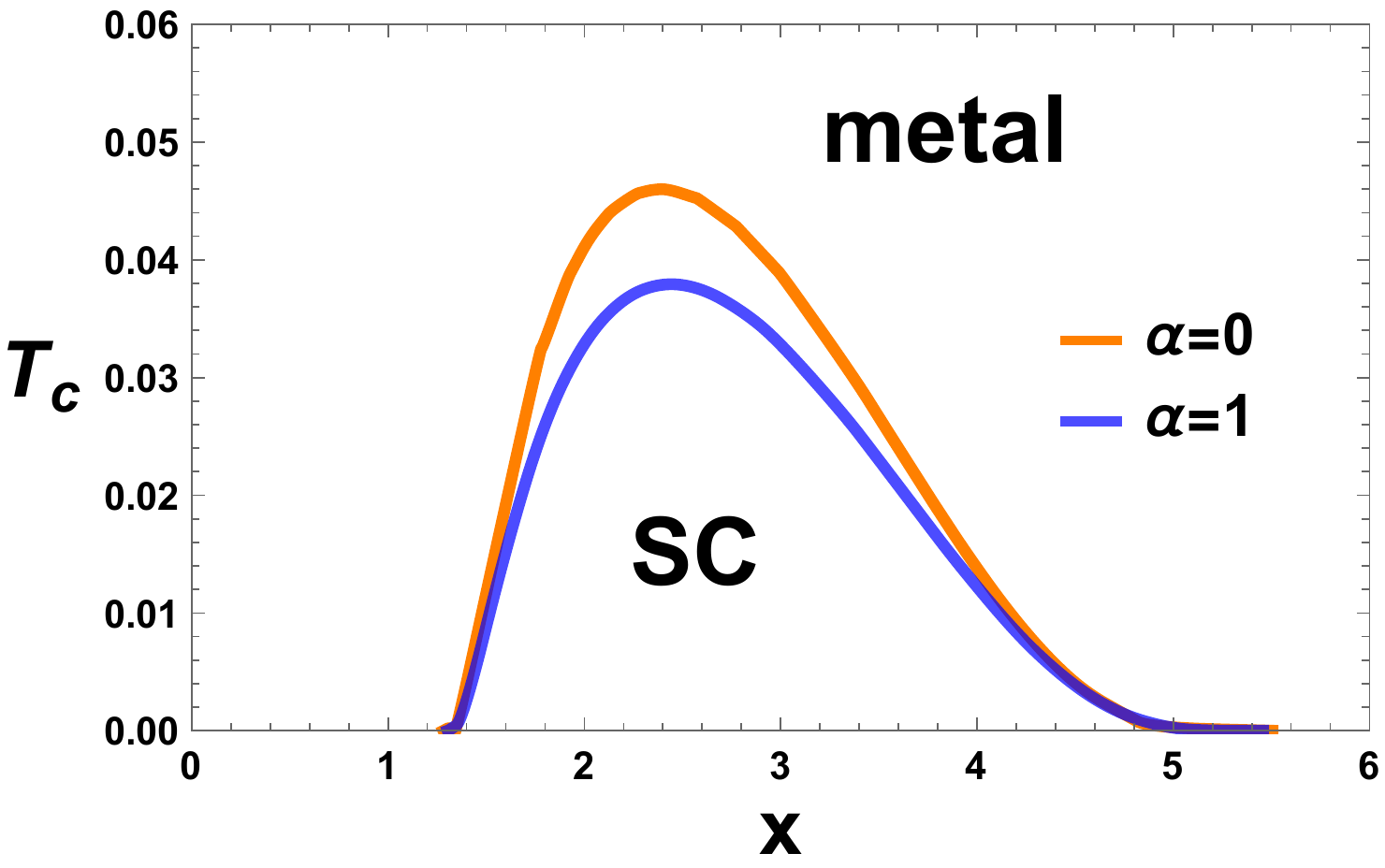}
\end{center}
\caption{Phase diagram in the $(T,{\bf x})$ plane for the model (\ref{Kiritsisparameters}) coupled to
the neutral scalars with the Lagrangian
(\ref{linearV}). We compare the case $\alpha=0$ of \cite{Kiritsis:2015hoa}, and
the system with broken translational symmetry, at $\alpha=1$.}
\label{fig:phasediagram}
\end{figure}

Now let us fix the value of $\alpha$ and plot the critical temperature
as a function of the doping parameter ${\bf x}$, see figure \ref{fig:phasediagram}.
The breaking of translation symmetry preserves the superconducting
dome structure exhibited by the model (\ref{Kiritsisparameters}), and merely diminishes a little
the critical temperature. 

\begin{figure}
\begin{center}
\includegraphics[width=.45\textwidth]{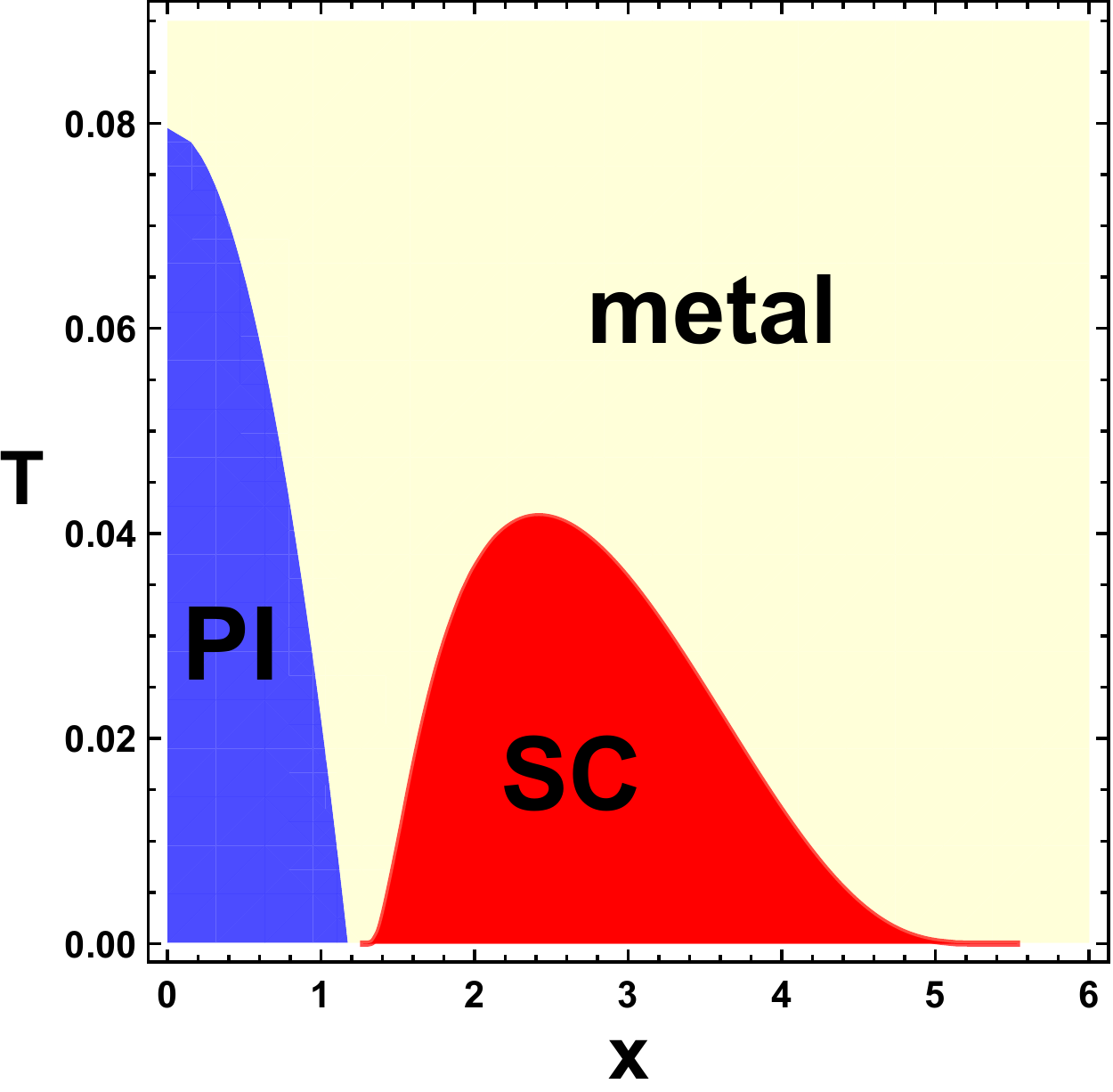}
\end{center}
\caption{Phase diagram in the $(T,{\bf x})$ plane for the model (\ref{Kiritsisparameters}) coupled to
the neutral scalars with the Lagrangian
(\ref{nonlinearV}). We fixed $\alpha=0.5$ and $m=1$.}
\label{fig:nonlinearphasediagram}
\end{figure}

Now let us consider the model with translational
symmetry broken by neutral scalars governed by the non-linear
Lagrangian (\ref{nonlinearV}).
We fix $\alpha=0.5$, $m=1$ and determine the critical temperature $T_c({\bf x})$.
In figure \ref{fig:nonlinearphasediagram} we combine this with the temperature $T_0({\bf x})$ of the
metal/pseudo-insulator phase transition (MIT), described in section \ref{sec:normalphase},
and obtain the full phase diagram of the system
with the superconducting phase enclosed inside a dome.

This means that even if momentum dissipation unfavores the SC phase it is still possible to achieve a SC dome-shaped region as in actual High-Tc superconductors and having a normal phase with a finite DC conductivity. This is the main result of our paper.

\section{Condensate}
\label{sec:condensate}

In the previous section we studied the instability of the normal phase (\ref{normalphase})-(\ref{normalphase2})
towards development of a non-trivial profile of the scalar $\chi(u)$.
Observing an instability at temperature $T=T_c$ on its own is not
sufficient for the conclusion that the system exhibits a phase transition at this point.
Indeed, the instability analyses relies on the assumption that the phase transition
is a continuous second-order phase transition. 
To determine whether this is actually the case, one should calculate behavior of the
order parameter as a function of temperature, and make sure the continuous critical
point is not shielded by a first-order phase transition.

In our case we need to solve numerically five background equations of motion for
the model (\ref{totalaction}). These equations are provided in appendix \ref{app:BackgroundEqs}.
The order parameter $\langle{\cal O}\rangle (T)$ is read off as the coefficient
of the sub-leading term in the near-boundary expansion of the $\chi(u)$.
The methodology of a numerical solution for the background is practically identical to the one
we performed recently in 
\cite{Baggioli:2015zoa}, which the interested reader is encourage to consult for the details.
The only new subtlety now is that we need to keep the doping parameter ${\bf x}$
fixed. With care this can be achieved, for example, using the {\tt FindRoot}
function in {\it Mathematica}, now applied to solve for the $A_t'(u_h)$, $B_t'(u_h)$
for each fixed $\chi(u_h)$, such that both the source $C_-$ of the field $\chi$
at the boundary vanishes, and the ${\bf x}$ is fixed.

We use scaling symmetry of the background
equations of motion to fix $u_h=1$.
We plot the condensate as a function of temperature in the linear model
(\ref{linearV}), for $\alpha =0$ (the case of \cite{Kiritsis:2015hoa}),
and $\alpha =1$; and in the non-linear model
(\ref{nonlinearV}), for $\alpha =0.5$, $m=1$.

\begin{figure}
\begin{center}
\includegraphics[width=.45\textwidth]{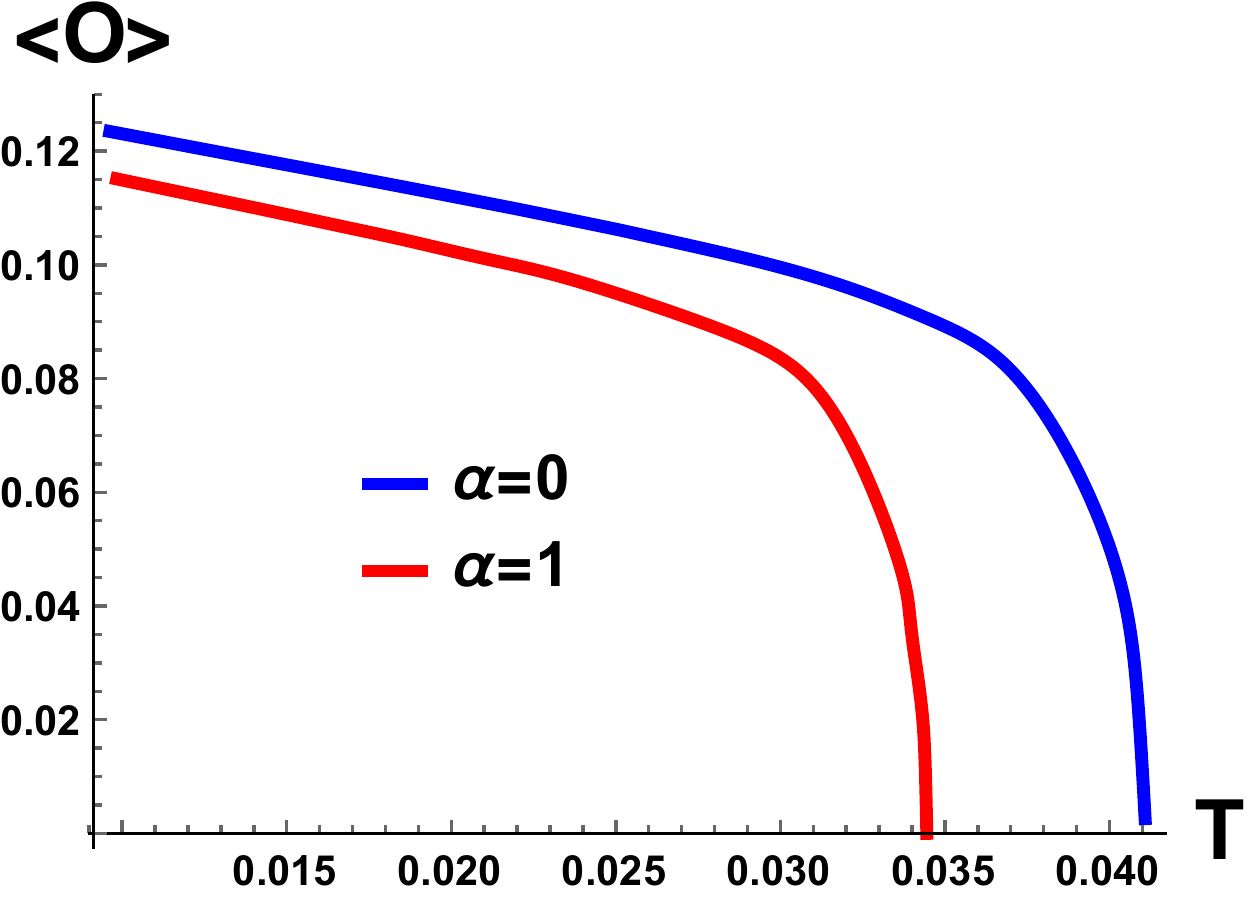}
\includegraphics[width=.45\textwidth]{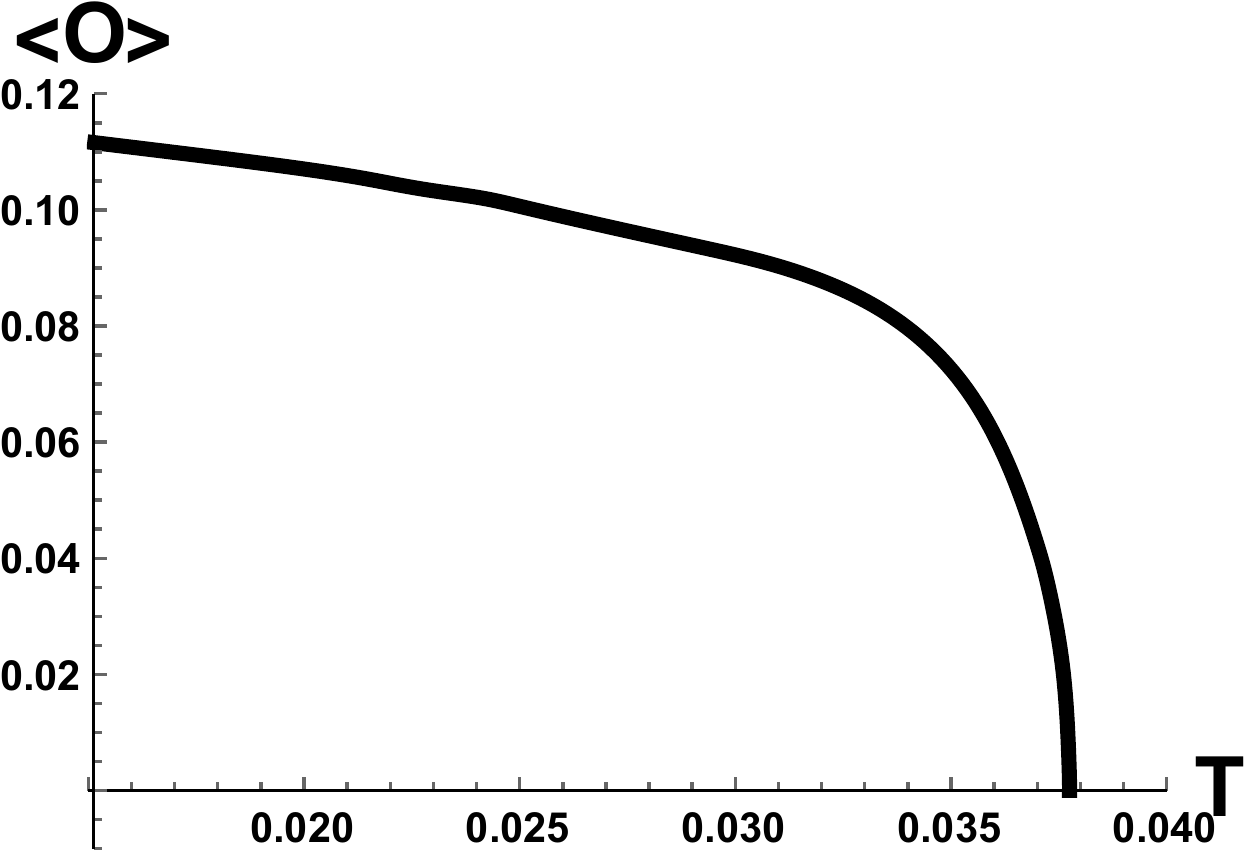}
\end{center}
\caption{Condensate for the model (\ref{Kiritsisparameters}), with the
doping fixed to ${\bf x}=2$, with broken translational symmetry.
{\bf Left:} The linear model (\ref{linearV}) with $\alpha=1$,
plotted next to the translationally-symmetric system $\alpha=0$.
{\bf Right:} The non-linear model (\ref{nonlinearV}) with $\alpha=0.5$, $m=1$.
The condensate is measured in units $\rho_A^{5/2}$, the temperature
is measured in units $\rho_A^{1/2}$. }
\label{fig:CondensateLinearV}
\end{figure}

We checked numerically the free energy of the system for the cases analyzed and we found that the SC phase whenever present is favoured. The interested reader can find details in \citep{Kiritsis:2015hoa} or \citep{Baggioli:2015zoa}.

\section{Discussion}
\label{sec:discussion}

In this paper we described a straightforward generalization of the holographic superconductor model
proposed in \cite{Kiritsis:2015hoa}. An important feature pointed out in \cite{Kiritsis:2015hoa}
reveals that introducing a non-trivial coupling between the order parameter
and the gauge fields one can achieve an enclosure of the superconducting phase
inside a dome-shaped region on the doping/temperature plane as in actual High-Tc superconductors.
We have expanded the model of \cite{Kiritsis:2015hoa} introducing a simple momentum dissipation mechanism through a sector
of neutral and massless scalars, breaking translational symmetry \citep{Andrade:2013gsa}. The conclusion is that in a generic
situation the superconducting dome of \cite{Kiritsis:2015hoa} survives the translational symmetry breaking and can be equipped with a normal phase featuring a finite DC conductivity.

In the case of a non-linear Lagrangian for the neutral scalars \cite{Baggioli:2014roa}
the normal phase can be further
split into two phases, distinguished by the sign of the first temperature derivative
of the DC conductivity. When this sign is negative, the system behaves like
a metal, when it is positive, it resembles an insulator. We pointed out
that for a generic choice of translational symmetry
breaking parameters an insulator occupies a compact region in the corner of the $({\bf x},T)$ plane.
The total resulting phase diagram exhibits three phases: metal, superconductor,
and pseudo-insulator.

The main result of this paper is to show that the SC dome-shaped region built in \cite{Kiritsis:2015hoa} can be completed with a simple momentum dissipation mechanism and embedded in a normal phase region featuring a finite DC conductivity. This represents a further step towards reproducing holographically the phase diagram for High-Tc superconductors.

It would be interesting to incorporate the translational symmetry breaking 
framework into the whole phase diagram, constructed in \cite{Kiritsis:2015hoa},
which also includes normal ferromagnetic and stripe/checkerboard phases.
It would also be interesting to calculate the AC conductivity and study
the collective excitation in the non-linear $V$ model, pointed out in \cite{Baggioli:2014roa}, and further
investigated in the holographic superconductor of \cite{Baggioli:2015zoa}. Because of the non trivial couplings between the various sectors and the two gauge fields we expect interesting features to appear.\\
Finally it would be great to generalize the model to account for a \textit{real} insulating normal state with $\sigma_{DC}(T=0)=0$. In order to do so one has to break the assumptions of \citep{Grozdanov:2015qia}; results in this direction are coming soon \cite{Baggioli:upcoming}.

\acknowledgments
M.B. would like to thank the University of Illinois and P.Phillips for the warm hospitality during the completion of this work. M.B. acknowledges support from MINECO under grant FPA2011-25948,  DURSI under grant 2014SGR1450 and Centro de Excelencia Severo Ochoa program, grant SEV-2012-0234 and is supported by a PIF grant from Universitat Autonoma de Barcelona UAB.
M.G. would like to acknowledge support from the Oehme Fellowship.

\appendix

\section{Background equations of motion}\label{app:BackgroundEqs}
The equations of motion following from the action (\ref{totalaction}) for the ansatz defined in (\ref{generalansatz}) are:
\begin{align}
&\left(u\,\chi ^{\prime 2}-2\tau '\right)f^2+2\,u\, e^\tau H\,(q_A\,A_t+q_B\,B_t)^2=0\,,\nonumber\\
&4\,u\,f'-(12+u^2\chi^{\prime 2})f-2\,u^2 e^\tau H\,\frac{(q_A\,A_t+q_B\,B_t)^2}{f(u)}+
12-2\,L^2\,(2\,m^2\,V+V_{int})\nonumber\\
&-e^\tau u^4\left(Z_{AB}\,A_t^{\prime 2}+B_t'\,(2\,Z_{AB} \, A_t'+Z_B\,B_t')\right)=0\,,\nonumber\\
&Z_A\,(2\,A_t''+\tau'\,A_t')+Z_{AB}\,(2\,B_t''+\tau'\,B_t')+2\,\chi' (\dot Z_A\,A_t'+\dot Z_{AB}\,B_t')
-4\,q_A\, H\,\frac{q_A \,A_t+q_B\, B_t }{u^2\,f}=0\,,\nonumber\\
&Z_B\,(2\,B_t''+\tau'\,B_t')+Z_{AB}\,(2\,A_t''+\tau'\,A_t')+2\chi' (\dot Z_B\,B_t'+\dot Z_{AB}\,A_t')
-4\,q_B\, H\,\frac{q_A \,A_t+q_B\, B_t }{u^2\,f}=0\,,\nonumber\\
&\chi''{+}\left(\frac{f'}{f}{-}\frac{2}{u}{-}\frac{\tau'}{2}\right)\chi'{-}\frac{L^2}{u^2\,f}\dot V_{int}
{+}\frac{e^\tau \,u^2}{2\,f}\left(\dot Z_A\,A_t^{\prime 2}{+}\dot Z_B\,B_t^{\prime 2}
{+}2\,\dot Z_{AB}\,A_t'\,B_t'\right)
{+}\frac{e^\tau \,\dot H}{f^2}\left(q_A\,A_t{+}q_B \,B_t\right)^2{=}0\,.\nonumber
\end{align}
Here dot stands for a derivative w.r.t. the scalar $\chi$, and prime stands for a derivative w.r.t. the radial coordinate $u$.


\begin{thebibliography}{1234567}
\bibitem{Hartnoll:2008vx} 
  S.~A.~Hartnoll, C.~P.~Herzog and G.~T.~Horowitz,
  ``Building a Holographic Superconductor,''
  Phys.\ Rev.\ Lett.\  {\bf 101}, 031601 (2008)
  [arXiv:0803.3295 [hep-th]].
\bibitem{Hartnoll:2008kx} 
  S.~A.~Hartnoll, C.~P.~Herzog and G.~T.~Horowitz,
  ``Holographic Superconductors,''
  JHEP {\bf 0812}, 015 (2008)
  [arXiv:0810.1563 [hep-th]].
\bibitem{Kiritsis:2015hoa} 
  E.~Kiritsis and L.~Li,
  ``Holographic Competition of Phases and Superconductivity,''
  arXiv:1510.00020 [cond-mat.str-el].
\bibitem{Vegh:2013sk} 
  D.~Vegh,
  ``Holography without translational symmetry,''
  arXiv:1301.0537 [hep-th].
\bibitem{Andrade:2013gsa} 
  T.~Andrade and B.~Withers,
  ``A simple holographic model of momentum relaxation,''
  JHEP {\bf 1405}, 101 (2014)
  [arXiv:1311.5157 [hep-th]].
\bibitem{Andrade:2014xca} 
  T.~Andrade and S.~A.~Gentle,
  ``Relaxed superconductors,''
  JHEP {\bf 1506}, 140 (2015)
  [arXiv:1412.6521 [hep-th]].
\bibitem{Kim:2015dna} 
  K.~Y.~Kim, K.~K.~Kim and M.~Park,
  ``A Simple Holographic Superconductor with Momentum Relaxation,''
  JHEP {\bf 1504}, 152 (2015)
  [arXiv:1501.00446 [hep-th]].
\bibitem{Baggioli:2015zoa} 
  M.~Baggioli and M.~Goykhman,
  ``Phases of holographic superconductors with broken translational symmetry,''
  JHEP {\bf 2015}, 35
  [arXiv:1504.05561 [hep-th]].
\bibitem{Baggioli:2014roa} 
  M.~Baggioli and O.~Pujolas,
  ``Electron-Phonon Interactions, Metal-Insulator Transitions, and Holographic Massive Gravity,''
  Phys.\ Rev.\ Lett.\  {\bf 114}, no. 25, 251602 (2015)
  [arXiv:1411.1003 [hep-th]].
\bibitem{Blake:2013bqa} 
  M.~Blake and D.~Tong,
  ``Universal Resistivity from Holographic Massive Gravity,''
  Phys.\ Rev.\ D {\bf 88}, no. 10, 106004 (2013)
  [arXiv:1308.4970 [hep-th]].
\bibitem{Donos:2014cya} 
  A.~Donos and J.~P.~Gauntlett,
  ``Thermoelectric DC conductivities from black hole horizons,''
  JHEP {\bf 1411}, 081 (2014)
  [arXiv:1406.4742 [hep-th]].
\bibitem{Grozdanov:2015qia} 
  S.~Grozdanov, A.~Lucas, S.~Sachdev and K.~Schalm,
  ``Absence of disorder-driven metal-insulator transitions in simple holographic models,''
  arXiv:1507.00003 [hep-th].
  \bibitem{Baggioli:upcoming} 
  M.~Baggioli and O.~Pujolas,
  ``A note on disorder-driven metal-insulator
transitions in holographic models,''
  to appear soon.

\end{thebibliography}
\end{document}